\documentclass[lettersize,journal]{IEEEtran}
\usepackage{amsmath,amsfonts}
\usepackage{algorithmic}
\usepackage{algorithm}
\usepackage{array}
\usepackage[caption=false,font=normalsize,labelfont=sf,textfont=sf]{subfig}
\usepackage{textcomp}
\usepackage{stfloats}
\usepackage{url}
\usepackage{verbatim}
\usepackage{graphicx}
\usepackage{cite}
\usepackage{float}
\usepackage{wrapfig}  
 
\begin{document}

\title{The first calibration
model for bluetooth angle of arrival: Enhancing positioning accuracy in
indoor environments}

\author{Ma'mon Saeed Alghananim$^*$,Yuxiang Feng,~\IEEEmembership{Member,~IEEE,}
        and Washington Yotto Ochieng 
        
\thanks{All authors are with the Centre for Transport Engineering and Modelling, Department of Civil and Environmental Engineering, Imperial College London, London SW7 2AZ, U.K.}% <-this % stops a space
\thanks{$^*$Corresponding author: Ma'mon Saeed Alghananim (e-mail: m.alghananim17@imperial.ac.uk)}}



\maketitle

\begin{abstract}
Internet of Things (IoT) applications are increasingly reliant on indoor positioning systems to deliver precise and reliable navigation in GNSS-denied environments, including urban areas, smart warehouses, hospitals, and underground or multi-level parking systems. Bluetooth Angle of Arrival (AoA) positioning offers cost-effective solutions with the potential to provide users with sub-meter position accuracy, which is crucial for applications such as underground navigation, firefighters, and robotic navigation. Bluetooth AoA positioning uses angles to determine the position of Bluetooth tags; these angles, measured in the anchor coordinate system, need to be transferred to the user’s coordinate system. This requires models or techniques to compute 3D rotation matrices between the anchor and user coordinate system. Until now, no model or technique has been developed to compute these rotation matrices. Therefore, the development of the AoA positioning model focuses on simulated scenarios. This paper introduces the first model, named the AoA calibration model, capable of estimating these rotation matrices, thereby facilitating the practical application of this technology. In addition, this paper tests the Bluetooth AoA calibration and positioning model on a real dataset and presents end-to-end functional architectures for AoA positioning. The results demonstrate that the proposed calibration model can estimate the 3D transformation rotation angles with a standard deviation better than 2.5 degrees. The findings also reveal that AoA positioning can achieve sub-meter accuracy in both static and kinematic modes, with accuracy significantly influenced by the distance to the anchors and the geometry factor.

\end{abstract}


\section{Introduction}
\IEEEPARstart{A}{s} Internet of Things (IoT) applications expand, Indoor Positioning Systems (IPS) play a critical role in enabling real-time location tracking and enhancing the functionality of connected devices. In recent years, there has been a growing interest in IPS due to the increasing demand for location-based services in indoor environments across a wide range of applications, such as pedestrian navigation, emergency services, robotic navigation, firefighting, indoor mapping, underground navigation, mining, and navigation for the visually impaired.

In the Positioning, Navigation, and Timing (PNT) landscape, IPS can serve as an alternative to GNSS in indoor environments where GNSS signals are obstructed and as an aided system or alternative for GNSS in complex environments where GNSS signals are affected by multipath interference. IPS can be based on one or more positioning models, including position fixing, dead reckoning, and computer vision. The primary technologies used for position fixing include, but are not limited to, Wi-Fi, Bluetooth, Infrared, RFID, and ZigBee. When comparing these technologies, it is important to consider the entire end-to-end positioning operation, including the required infrastructure and its availability (such as the number of nodes and their geometry), cost, and system requirements. Accuracy, for instance, is influenced by several factors, such as measurement types and their accuracy, positioning methodology, number of measurements, positioning algorithm, geometry, anchors density, and operating environment. Therefore, a comparison of these technologies shall take these factors into account.

Generally, Wi-Fi and ZigBee positioning can provide accuracy within 1-5 metres \cite{ref1}, \cite{ref2}, \cite{ref3}, \cite{ref4}. RFID-based IPS can provide meter-level positional accuracy in active mode using Received Signal Strength Indicator (RSSI), and decimetre accuracy level in passive mode; however, it has a limited range of approximately 2 meters \cite{ref5}. UWB is capable of providing centimetre accuracy level \cite{ref6}, \cite{ref7}. However, it comes with high infrastructure costs.

Bluetooth-based IPS has the potential to provide sub-meter to meter-level accuracy, depending on the positioning algorithm, and offers a cost-effective alternative with significantly lower expenses compared to UWB. Bluetooth positioning can be classified based on measurements into RSSI-based positioning and/or Angle of Arrival (AoA) based positioning. RSSI-based positioning provides accuracy within 2-10 meters using one or more positioning models: trilateration \cite{ref8}, \cite{ref9}, multilateration \cite{ref10}, proximity detection \cite{ref11}, \cite{ref12} and Received Signal Strength Mapping (RSSM) \cite{ref9}, \cite{ref13}, \cite{ref14}. AoA-based positioning, on the other hand, is a promising model that can provide angle measurements with a mean absolute error of 5 degrees \cite{ref15}, which can be projected in the positioning domain to achieve sub-meter accuracy.

Most previous studies have focused on estimating AoA measurements, employing methods such as the Multiple Signal Classification (MUSIC) \cite{ref16}, \cite{ref17}, and I/Q density-based angle of arrival estimation \cite{ref18}. In the development of a positioning algorithm using AoA measurements, \cite{ref19} introduced a database-based AoA positioning model that relies on a database containing computed position information within the grid and its AoA measurements. This approach is quite expensive in terms of implementation, as it requires a large database to be stored in the system. Conversely, few studies have focused on developing AoA positioning models using non-database approaches \cite{ref20}, \cite{ref21}, \cite{ref22}, however, these models have only been tested in simulation environments.

In real-world implementation, the orientation of the AoA anchors around the x, y, and z axes must be computed and modelled within the positioning functional model. In other words, since AoA utilizes angles to compute the position, and these angles are computed in the Bluetooth anchor coordinate system, they must be transferred to the user's coordinate system. This requires transformation matrices that model the AoA anchor's orientation in space. To the best of our knowledge, no model or technique has been developed to compute the anchor’s orientation or to define the rotation matrices between the anchors and user’s coordinate systems. This gap makes testing models in real, non-simulated scenarios very complex. Even the database-based AoA positioning model, which was tested with real data, did not model the anchor’s orientation. 

To address the above-mentioned limitations, this paper makes the following contributions:
\begin{itemize}
    \item It introduces, for the first time, a calibration model capable of estimating the orientation of AoA anchors. This model defines the rotation matrices between the user and anchor coordinate systems within the positioning algorithm. This model is anticipated to enable the reliable implementation of AoA Bluetooth technology. 
    \item It presents AoA Bluetooth positioning models, which incorporate the rotation matrices derived from the calibration model. 
    \item It presents functional architectures for the complete end-to-end process of AoA Bluetooth positioning, providing a comprehensive framework that has not been presented in previous studies.
    \item it tests the developed calibration and positioning algorithms in real dataset. 
\end{itemize}

This paper is organized as follows: Section~\ref{sec:FUNCTIONAL ARCHITECTURE} presents the functional architectures for the complete end-to-end process of AoA Bluetooth positioning. Section~\ref{sec:CALIBRATION MODEL} introduces the proposed calibration model along with its mathematical framework. Section~\ref{sec:ANGLE OF ARRIVAL POSITIONING MODEL} explains the mathematical model of AoA positioning. Section~\ref{sec:RESULTS} highlights the results obtained from testing the calibration and positioning models on real datasets. Finally, Section~\ref{sec:CONCLUSION AND FUTURE WORK} concludes the work and provides insights into potential future research directions.
\section{Functional Architecture}
\label{sec:FUNCTIONAL ARCHITECTURE}
\begin{figure*}[!t]
    \centering
	\includegraphics[width=0.88\textwidth]{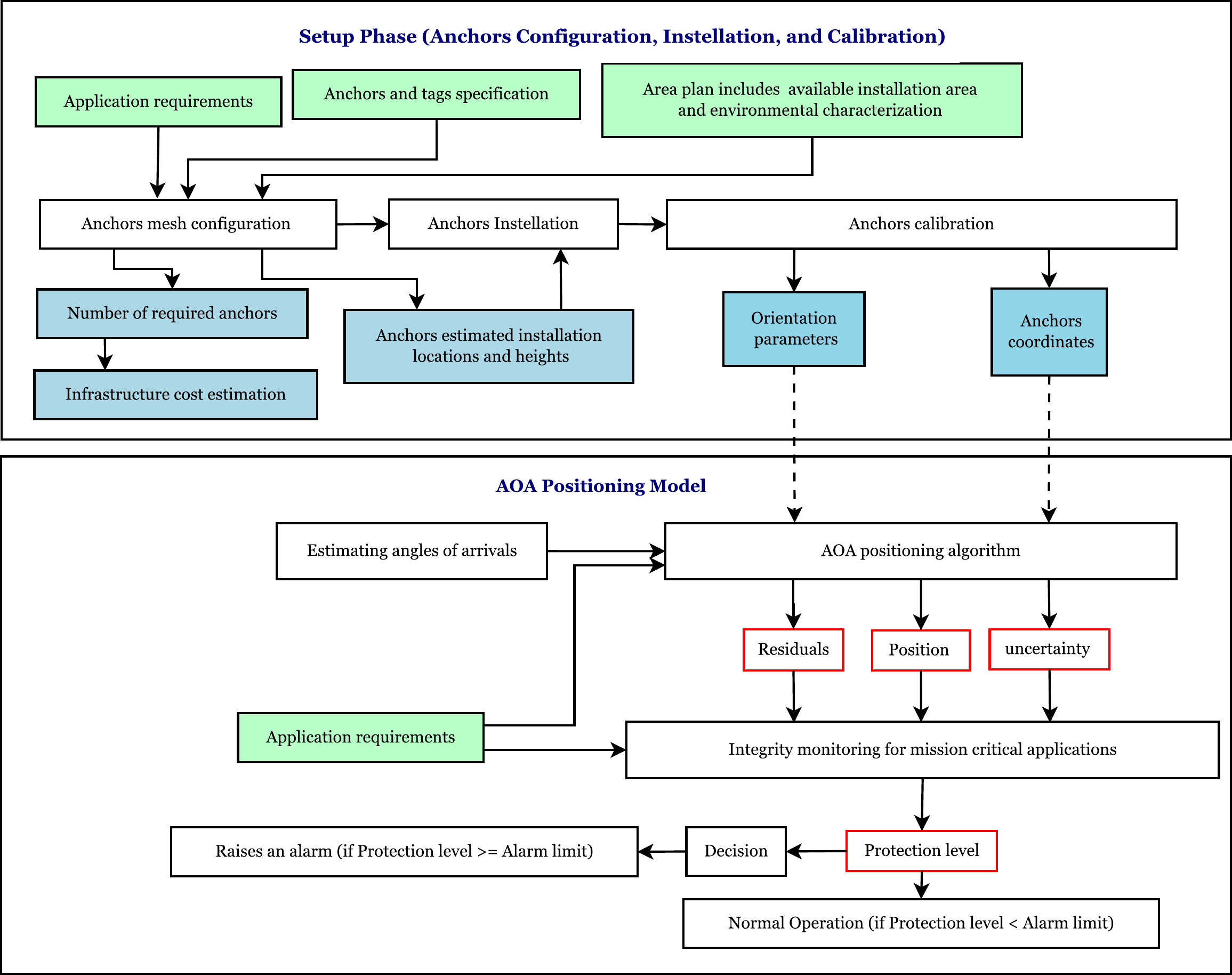}
	\caption{High-level functional architecture for the end-to-end Bluetooth AoA positioning, including the setup phase and positioning model.} 
	\label{fig.1}
\end{figure*}

Fig.~\ref{fig.1} shows the functional architecture for Bluetooth AoA positioning and calibration model, divided into two main phases: the setup phase and the AoA positioning phase. The following subsections discuss these phases in detail.

\subsection{Setup phase}
The setup phase (anchor configuration, installation, and calibration) aims to determine the optimal number and locations of Bluetooth anchors. The setup phase comprises three stages: anchor mesh configuration, anchor installation, and anchor calibration.

In the anchor mesh configuration stage, a configuration model identifies the required number of anchors and their optimal locations, orientation, and height. This is can be identified based on several factors, including, but not limited to, application requirements, anchor and tag specifications, environmental, positioning model/algorithm, and available installation sites within the area. The configuration model can also be used to estimate the infrastructure costs needed to meet the applications requirements. For indoor positioning technologies (e.g., UWB, RFID, and Wi-Fi), such models could be used to select the candidates technology for a specific area, considering performance requirements and infrastructure costs. However, to our knowledge, such a model has not yet been developed.

\begin{figure*}[!t]
    \centering
	\includegraphics[width=0.88\textwidth]{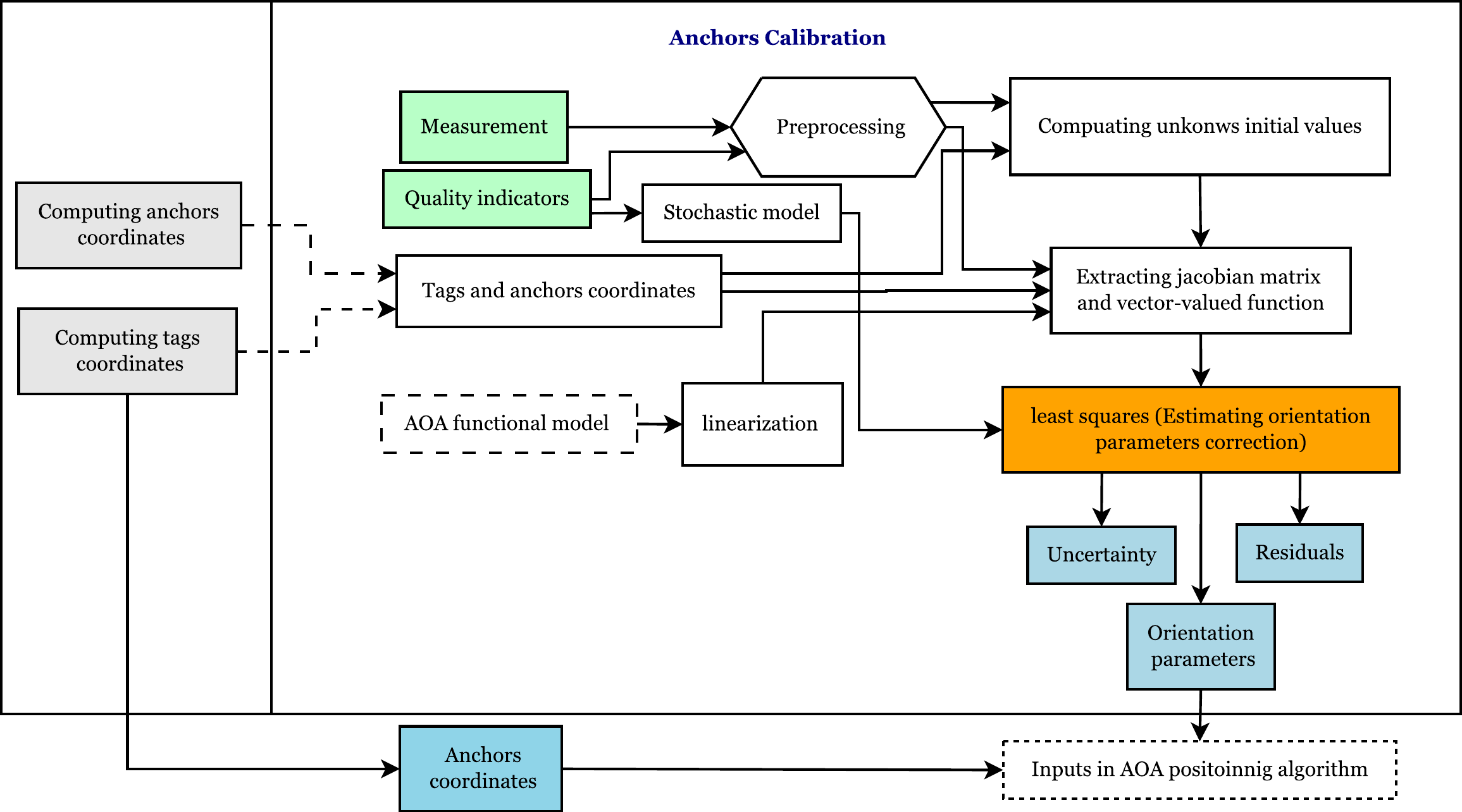}
	\caption{The functional architecture of the calibration model.} 
	\label{fig.2}
\end{figure*}

The anchor installation process involves placing Bluetooth anchors at their locations, heights, and orientations as defined during the anchor mesh configuration stage. Given the difficulty of precisely setting the anchors' orientations during installation, an anchor calibration process can be applied afterward to accurately estimates the anchors' coordinates and orientations. This paper introduces a novel calibration model utilizing the AoA position, with the rotation angle as an unknown variable and the tags' coordinates as known values. Detailed discussion of this model is provided in Section~\ref{sec:CALIBRATION MODEL}. The functional architecture of the proposed calibration model is illustrated in Fig.~\ref{fig.2}, which includes five main stages: pre-processing, computing unknown initial values, forming a stochastic model, extracting the Jacobian and observation matrix, and estimating the orientation parameters.

The pre-processing is intended to detect outliers using quality indicators (e.g., RSSI values) or variations in measurements during the observation session. Initial values for anchors orientation, which will be used in the nonlinear least squares process, can be estimated using the minimum required observations to estimate rotation parameters. Stochastic models can be used in cases where a weighted least squares approach is employed. Estimating the orientation parameters through nonlinear least squares, which is discussed in detail in Section~\ref{sec:CALIBRATION MODEL}, allows for estimating the anchor's orientation and its associated uncertainty. Following the least squares estimation, an additional quality-checking stage can be employed using a test statistic model to accept or reject the solution.

\subsection{The AOA positioning model}
\begin{figure*}[!t]
    \centering
	\includegraphics[width=0.88\textwidth]{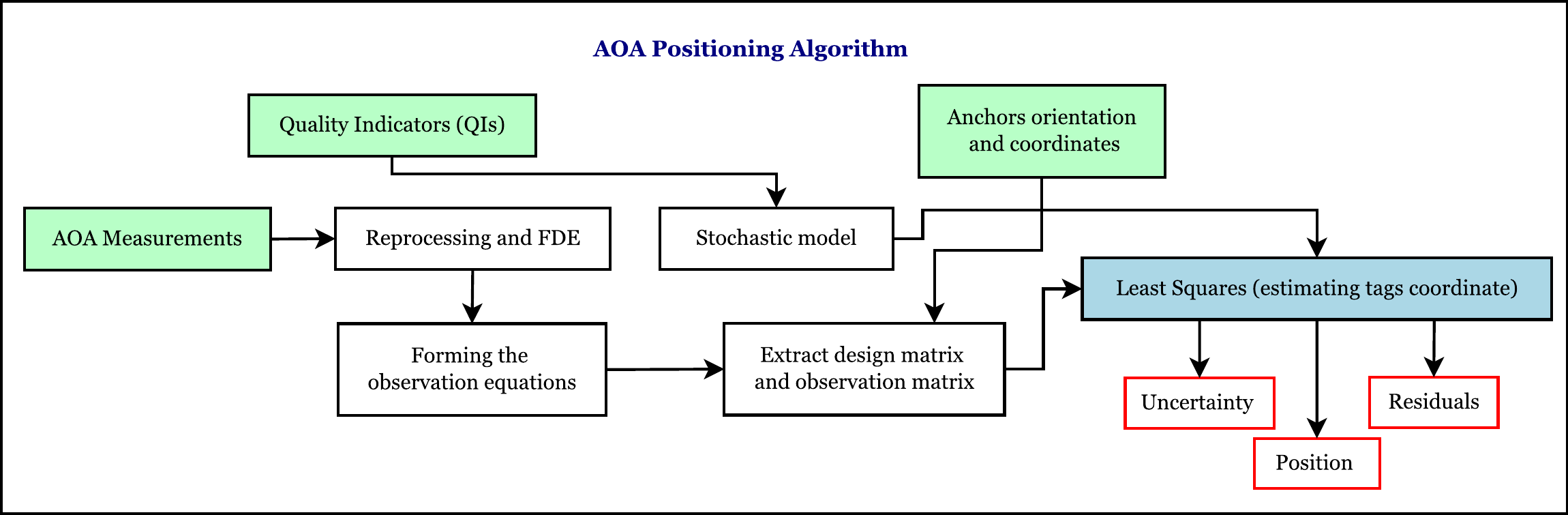}
	\caption{Functional architecture for Bluetooth AOA positioning.} 
	\label{fig.3}
\end{figure*}

The AoA positioning model aim to estimate tag positions in either real-time or post-processing mode using AoA measurements. Fig.~\ref{fig.3} presents the AoA positioning architecture, which includes six stages: pre-processing, forming the observation equation, extracting the design matrix and observation equation, forming the stochastic model, estimating tag positions using least squares, and integrity monitoring. The integrity monitory layer includes test statistics, Fault Detection and Exclusion (FDE), and protection level computation.

During the pre-processing stage, outlier detection and exclusion can be applied to enhance positioning performance by identifying outliers using quality indicators, such as RSSI. In the stochastic modelling stage, a weighted solution may be implemented using also the quality indicators. This paper proposes a non-weighted AoA model includes a mathematical framework, extraction of the design matrix and observation equation, and utilisation of least squares estimation, all of which are discussed in detail in Section~\ref{sec:ANGLE OF ARRIVAL POSITIONING MODEL}. 

The integrity monitoring layer is crucial for mission-critical applications, such as robotic navigation and navigation for visually impaired individuals, where accidents could lead to injury or death. In these applications, integrity monitoring aims to detect and exclude faulty measurements, overbound any assumptions made in the positioning algorithm, and compute the protection level. One of the key assumptions in the positioning algorithm is that the error distribution follows a Gaussian distribution. This assumption shall be overbounded to ensure safety using the Maximum Non-Bounded Difference (MnBD) method \cite{ref23}.

\section{Calibration Model}
\label{sec:CALIBRATION MODEL}

The proposed calibration model in this paper is based on the AOA positioning model, where the unknown orientation parameters (rotation about axes) are defined as free parameters, and the tag coordinates are defined as known values. Put differently, the calibration model based on shifting the unknown parameters in AOA positioning model to compute of the anchor's orientation.

After installing the Bluetooth anchor, tags need to be distributed around the Bluetooth anchor with strong geometry. The first step of the calibration model is to compute the anchors' coordinates $(x_0, y_0, z_0)$ and tags' coordinates $(x_{t_n}, y_{t_n}, z_{t_n})$, using any precise positioning techniques (e.g., classical surveying). Then, the unit vector between the tag and the anchor with reference to the anchor coordinate system $(u_X, u_y, u_z)$ and the unit vector with reference to the user’s coordinate system $(v_x, v_y, v_z)$ can be derived as follows:

\begin{equation}
\label{eqn_ux}
u_x = \cos(\alpha_n) \cos(90^\circ - \gamma_n) 
\end{equation}

\begin{equation}
\label{eqn_uy}
u_y = \cos(\alpha_n) \sin(90^\circ - \gamma_n) 
\end{equation}

\begin{equation}
\label{eqn_uz}
u_z = \sin(\alpha_n)
\end{equation}

\begin{equation}
\label{eqn_vx}
v_x = \frac{x_{t_n} - x_0}{\rho}
\end{equation}

\begin{equation}
\label{eqn_vy}
v_y = \frac{y_{t_n} - y_0}{\rho}
\end{equation}

\begin{equation}
\label{eqn_vz}
v_z = \frac{z_{t_n} - z_0}{\rho}
\end{equation}

where: $\alpha_n$ and $\gamma_n$ are the observed azimuth and elevation angles, respectively. $\rho$ represents the range between tags and anchors, which is given by:

\begin{equation}
\label{eqn_rho}
\rho = \sqrt{(x_t - x_0)^2 + (y_t - y_0)^2 + (z_t - z_0)^2}
\end{equation}

\begin{figure}[b]
    \centering
	\includegraphics[width=0.95\columnwidth]{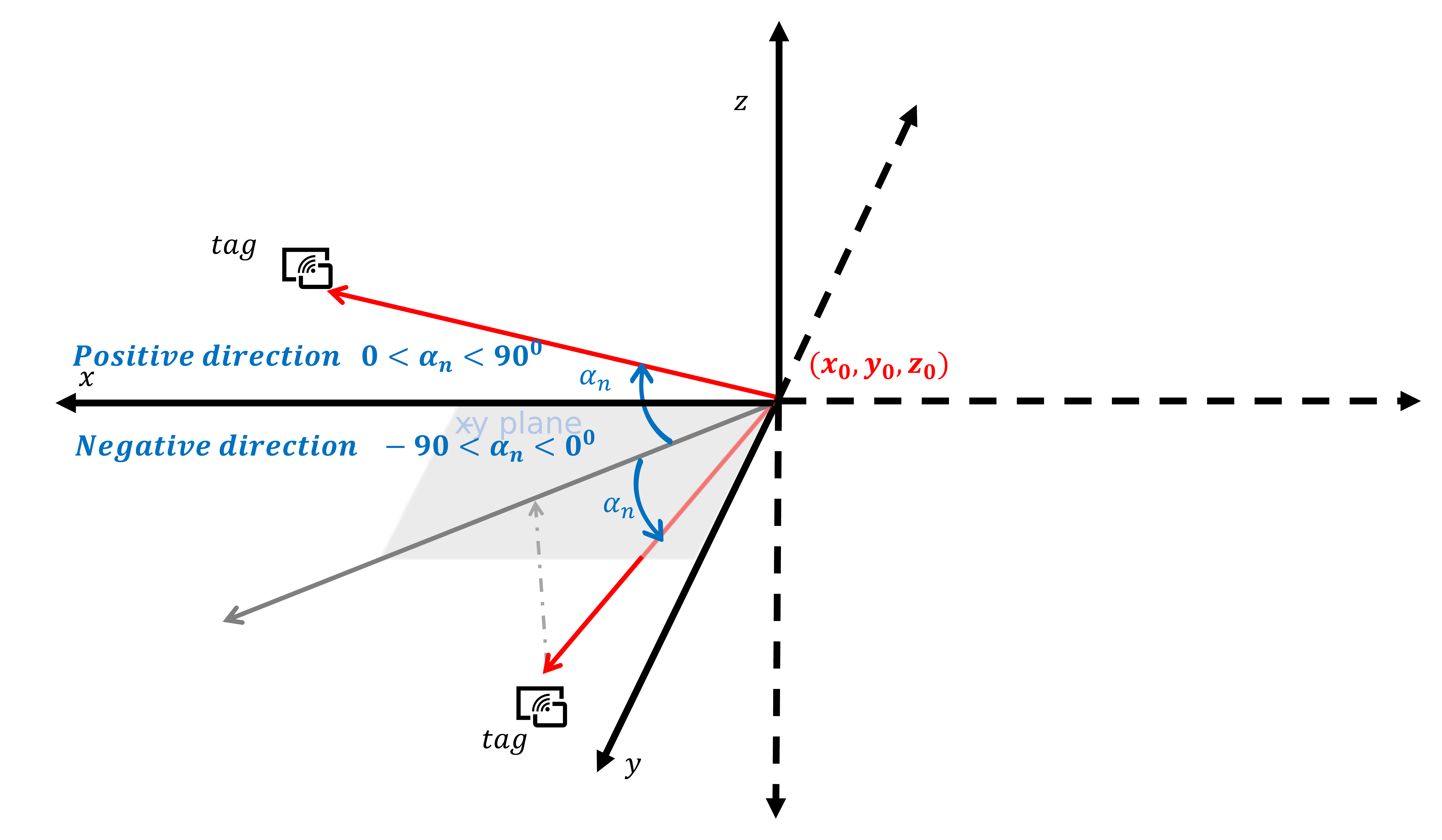}
	\caption{Illustration of elevation angle definitions relative to the Bluetooth anchors coordinate system.} 
	\label{fig.4}
    
\end{figure}

\begin{figure}[t]
    \centering
	\includegraphics[width=0.95\columnwidth]{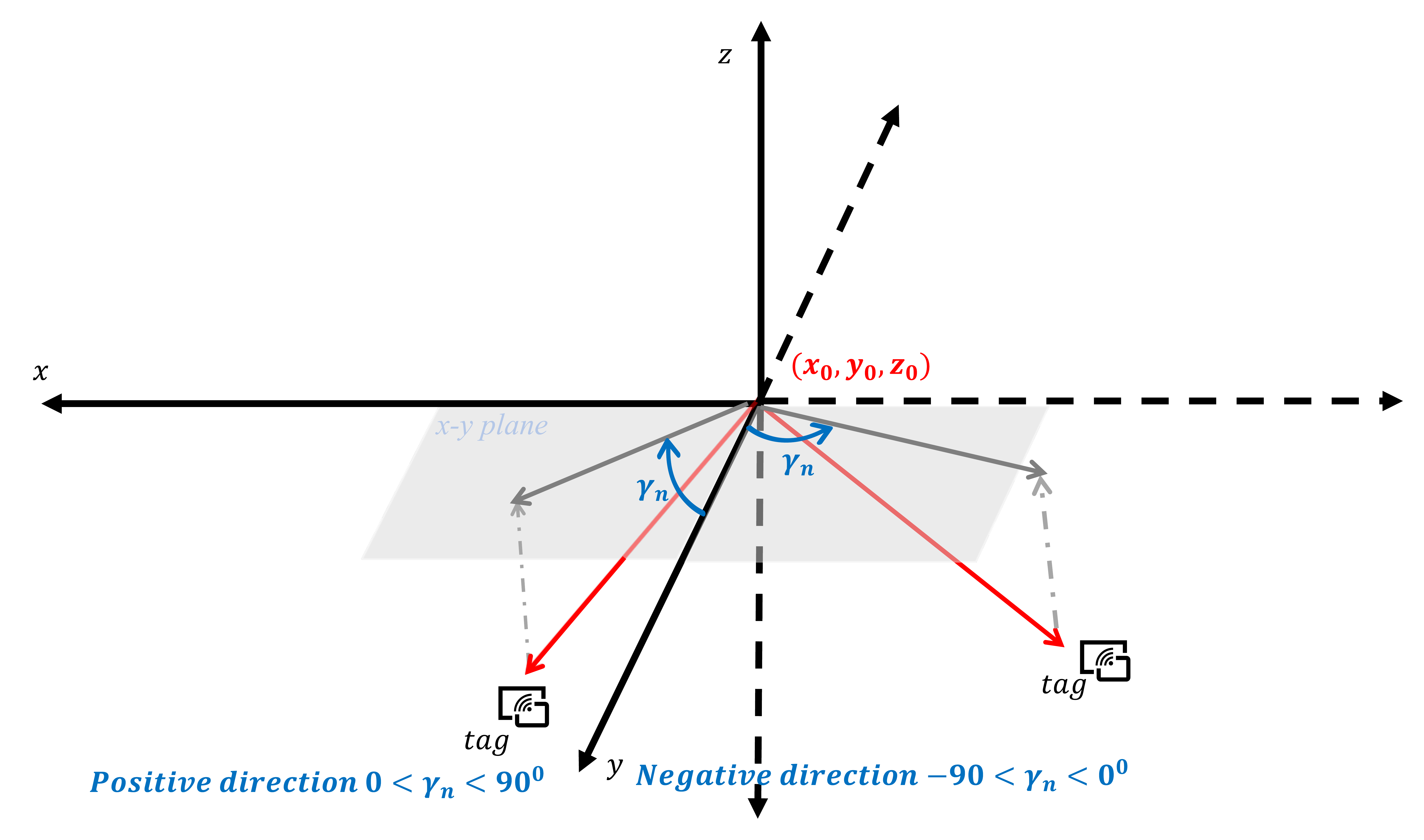}
	\caption{Illustration of azimuth angle definitions relative to the Bluetooth anchor’s coordinate system.} 
	\label{fig.5}
\end{figure}

In this paper, the coordinate system at the anchor level is established as a right-hand system. The elevation angle is measured from the horizontal XY plane, with values ranging from $0^\circ$ to $90^\circ$ in the upward direction and from $-90^\circ$ to $0^\circ$ in the downward direction, as shown in  Fig.~\ref{fig.4}. The azimuth angle is defined relative to the Y-axis and the projection of the line connecting the tag and the anchors origin. It ranges from $0^\circ$ to $90^\circ$ when the projection falls between the Y-axis and the positive X-axis, and from $-90^\circ$ to $0^\circ$ when the projection lies between the Y-axis and the negative X-axis, as illustrated in  Fig.~\ref{fig.5}.  


The functional model for the calibration model can then be defined as: 

\begin{equation}
\label{eqn_F}
F(\psi, \theta, \phi) = R_z(\psi) R_y(\theta) R_x(\psi) u - v = 0
\end{equation}

where: $R_z(\psi)$, $R_y(\theta)$, and $R_x(\phi)$ represent the 3D rotation matrices around the $z$, $y$, and $x$ axes, respectively, and are defined as:

\begin{equation}
\label{eqn_Rz}
R_z(\psi) = 
\begin{bmatrix}
\cos(\psi) & -\sin(\psi) & 0 \\
\sin(\psi) & \cos(\psi) & 0 \\
0 & 0 & 1
\end{bmatrix}
\end{equation}

\begin{equation}
\label{eqn_Ry}
R_y(\theta) = 
\begin{bmatrix}
\cos(\theta) & 0 & \sin(\theta) \\
0 & 1 & 0 \\
-\sin(\theta) & 0 & \cos(\theta)
\end{bmatrix}
\end{equation}

\begin{equation}
\label{eqn_Rx}
R_x(\phi) = 
\begin{bmatrix}
1 & 0 & 0 \\
0 & \cos(\phi) & -\sin(\phi) \\
0 & \sin(\phi) & \cos(\phi)
\end{bmatrix}
\end{equation}

From Equation 8, the following can be derived:
\begin{equation}
\label{eqn_fx_fy_fz}
\begin{bmatrix}
F_x \\ F_y \\ F_z
\end{bmatrix}
=
R_z(\psi) R_y(\theta) R_x(\psi)
\begin{bmatrix}
u_x \\ u_y \\ u_z
\end{bmatrix}
-
\begin{bmatrix}
v_x \\ v_y \\ v_z
\end{bmatrix}
= 0
\end{equation}
\begin{equation}
\label{eqn_vx_vy_vz}
\begin{bmatrix}
v_x \\ v_y \\ v_z
\end{bmatrix}
=
R_z(\phi) R_y(\theta) R_x(\psi)
\begin{bmatrix}
u_x \\ u_y \\ u_z
\end{bmatrix}
\end{equation}

The Jacobian matrix (J) and the observation matrix (B) can be extracted from the linearised from of equation (13). Then, the least squares solution is given by:
\begin{equation}
\label{eqn_H}
H = \Delta + H_0 = (J^t J)^{-1} J^t B + H_0
\end{equation}

$H$ denotes the updated estimate of the 3D rotation angles after applying the least squares solution, and $H_0$ represents the initial estimate of these angles. This solution is an iterative one and the results can be achieved when the correction ($\Delta$) becomes lower than a defined threshold, which can be defined based on the required computation precision. The iterative value is updated as follow:

\begin{equation}
\label{eqn_H0_condition}
H_0 = H \quad \text{if } \Delta > \text{threshold}
\end{equation}

\begin{equation}
\label{eqn_H_resolved}
H \text{ resolved if } \Delta < \text{threshold}
\end{equation}

The Mean Squared Error (MSE) and the estimate of the variance-covariance matrix ($\Sigma$) for the solution are given by:

\begin{equation}
\label{eqn_MSE}
\text{MSE} = \frac{(B - JH)^T \cdot (B - JH)}{m - n}
\end{equation}

\begin{equation}
\label{eqn_Sigma}
\Sigma = \left( J^T J \right)^{-1} \cdot \text{MSE}
\end{equation}

where: $m$ and $n$ represent the number of observations and unknowns, respectively.

\section{Angle of Arrival Positioning Model}
\label{sec:ANGLE OF ARRIVAL POSITIONING MODEL}

To compute the Bluetooth tag positions using the AoA model, the computation process can be summarized in the following steps:
\begin{itemize}
\item{} Create a unit vector between the tag and the anchors in the anchor coordinate system \((u_x, u_y, u_z)\), as given in Equations (1-3).  
\item{} Convert the created unit vectors from the anchor coordinate system to the user’s coordinate system \((v_x, v_y, v_z)\). This conversion requires the pre-defined orientation of the anchors in space \((\phi_0, \theta_0, \psi_0)\), computed in the anchor's calibration model. The transformation is given by:
\begin{equation}
\begin{bmatrix}
v_x \\ 
v_y \\ 
v_z
\end{bmatrix}
=
R_z(\phi) R_y(\theta) R_x(\psi)
\begin{bmatrix}
u_x \\ 
u_y \\ 
u_z
\end{bmatrix}
\end{equation}

\item{} Form the function model based on defining lines between known anchor coordinates \((x_0, y_0, z_0)\) and the unit vector. The functional models \((F_1, F_2)\) are derived as follows: 
\begin{equation}
\frac{x - x_0}{v_x} = \frac{y - y_0}{v_y} = \frac{z - z_0}{v_z}
\end{equation}
\begin{equation}
F_1 = (y - y_0) v_x - v_y (x - x_0) = 0
\end{equation}
\begin{equation}
F_2 = (x - x_0) v_z - v_x (z - z_0) = 0
\end{equation}

\item{} Extract the design matrix \(A\) and the observation matrix \(B\) from Equations (21, 22), the following is obtained: 
\begin{equation}
A =
\begin{bmatrix}
-v_{y_1} & v_{x_1} & 0 \\
v_{z_1} & 0 & -v_{x_1} \\
\vdots & \vdots & \vdots \\
-v_{y_n} & v_{x_n} & 0 \\
v_{z_n} & 0 & -v_{x_n}
\end{bmatrix}_{m \times 3}
\end{equation}
\begin{equation}
B =
\begin{bmatrix}
y_{0_1} v_{x_1} - x_{0_1} v_{y_1} \\
x_{0_1} v_{z_1} - v_{x_1} z_{0_1} \\
\vdots \\
y_{0_n} v_{x_n} - x_{0_n} v_{y_n} \\
x_{0_n} v_{z_n} - v_{x_n} z_{0_n}
\end{bmatrix}_{m \times 1}
\end{equation}

\item{} Implement an optimization technique to estimate the position by identifying the best intersection point in space. Using the least-squares method, the estimated position \((x, y, z)\), Mean Squared Error (MSE), and variance-covariance matrix are given by:
\begin{equation}
\begin{bmatrix}
x \\ 
y \\ 
z
\end{bmatrix}
=
\left( A^t A \right)^{-1} A^t B
\end{equation}
\begin{equation}
\text{MSE} = \frac{(B - Ax)^T \cdot (B - Ax)}{m - n}
\end{equation}
\begin{equation}
\Sigma = \left( A^T A \right)^{-1} \cdot \text{MSE}
\end{equation}

\end{itemize}

\section{Results}
\label{sec:RESULTS}

\begin{figure*}[!ht]
    \centering
	\includegraphics[width=0.9\textwidth]{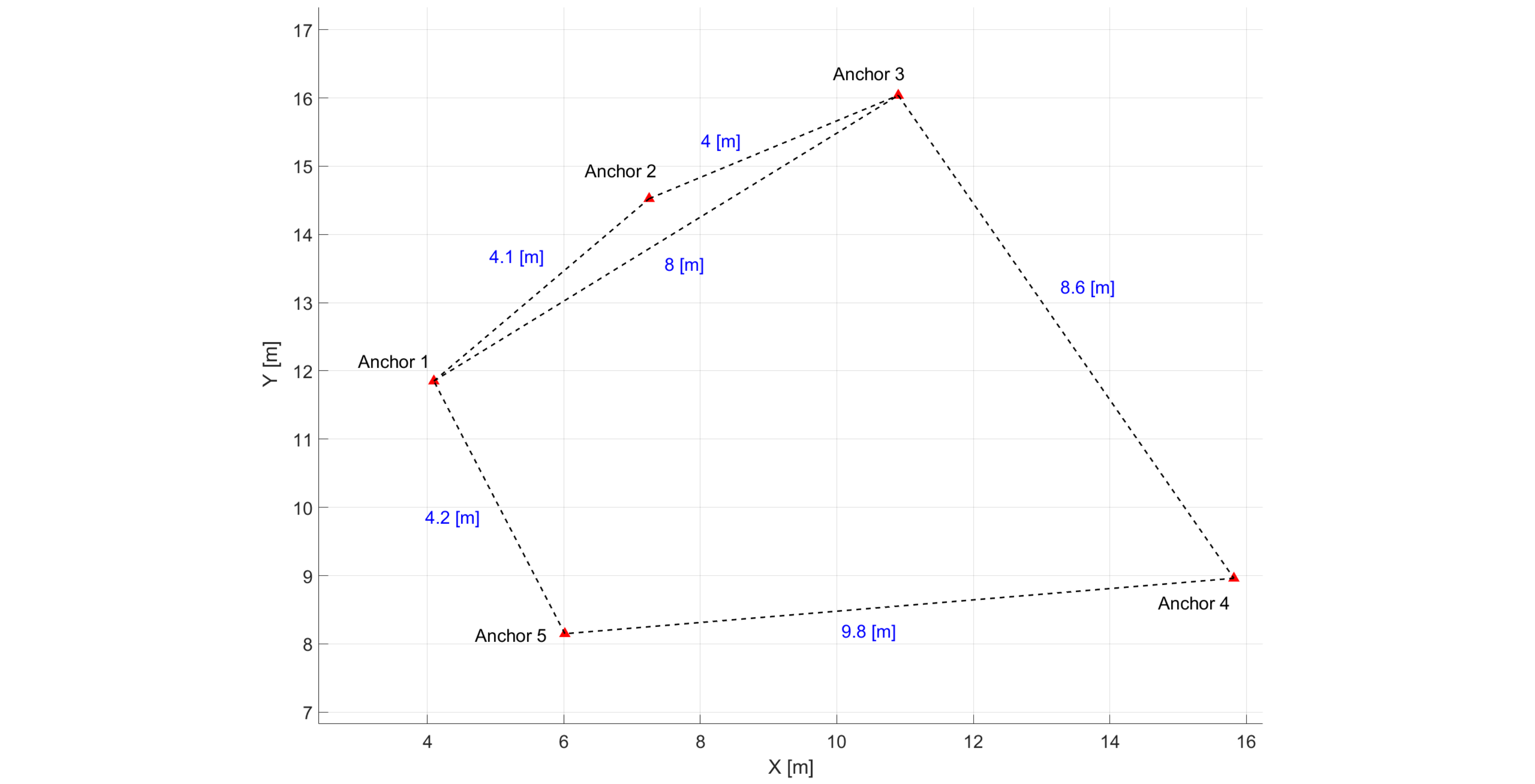}
	\caption{Anchors setup used in the experiments.} %\todo{replace with svg / no raster images}
	\label{fig.6}
\end{figure*}

The experiments were conducted in one of the Imperial College classrooms, covering approximately \(55 \, \text{m}^2\). Ublox’s XPLR-AOA-2 anchors and tags were utilized in the experiments, including five anchors as shown in ~\ref{fig.6}.

\begin{figure}[H]
    \centering
	\includegraphics[width=0.8\columnwidth]{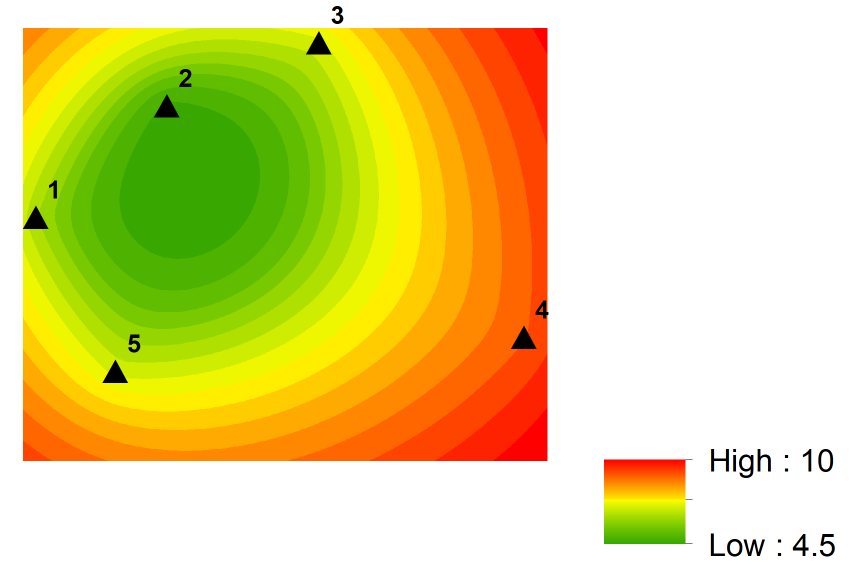}
	\caption{Average distance (in meters) between the anchors and various points in the space.} 
	\label{fig.7}
\end{figure}

The locations of the anchors were selected to maximize coverage within the room, taking into account the available installation sites. Four anchors were installed near corners of the room to provide good overall geometry for the positioning solution. Meanwhile, one of the anchors (Anchor 2) was installed between anchors 1 and 5 to create an area with a high anchors density. Fig.~\ref{fig.7} presents the average distance between the anchors and various points in the space, showing that the area around Point 2 has a lower average distance to the anchors compared to other areas. 

Given that the distance to the anchors is one of the indicators of positioning accuracy, it is expected that areas with a low average distance to the anchors would have higher positioning accuracy than others. Note, however, that this is not always the case, as accuracy also depends on other factors, such as geometry, number of anchors, calibration parameter accuracy, and the environment.

The rotation angles of the anchors were estimated using the proposed calibration approach. Ten tags were used for each anchor during the calibration process, with data collected over a 1-minute period. In the calibration model, only tags with a measurement standard deviation of \(3^\circ\) or less during the collection period were used. The calibration results, summarized in Table~\ref{Table.1}, show that the number of selected tags ranged from 6 to 8. The results indicate that the estimated standard deviations from the least-squares method for the three rotation angles—rotation around the x-axis (\(R_x\)), y-axis (\(R_y\)), and z-axis (\(R_z\))—fall between \(0.94^\circ\) and \(2.49^\circ\).
 
\begin{table*}[!t]
\caption{Calibration Results for Anchor Rotation Angles and Selected Tags}
\label{Table.1}
\centering
\footnotesize
\begin{tabular}{|c|c|c|c|c|c|c|c|}
\hline
ID & $R_x$ & $R_y$ & $R_z$ & $R_x$ std & $R_y$ std & $R_z$ std & Number of Selected Tags \\
\hline
1  & -42.4 & -0.4  & 249.5 & 0.9       & 1.3       & 1.8       & 6                      \\
2  & -46.1 & 0.1   & 193.0 & 1.0       & 2.4       & 2.5       & 8                      \\
3  & -48.0 & 1.8   & 173.1 & 0.4       & 1.2       & 1.3       & 6                      \\
4  & -15.2 & 16.2  & 64.3  & 1.1       & 1.8       & 1.9       & 8                      \\
5  & -51.6 & 2.1   & 295.0 & 1.5       & 1.4       & 2.2       & 6                      \\
\hline
\end{tabular}
\end{table*}

\begin{figure*}[!t]
    \centering
	\includegraphics[width=0.8\textwidth]{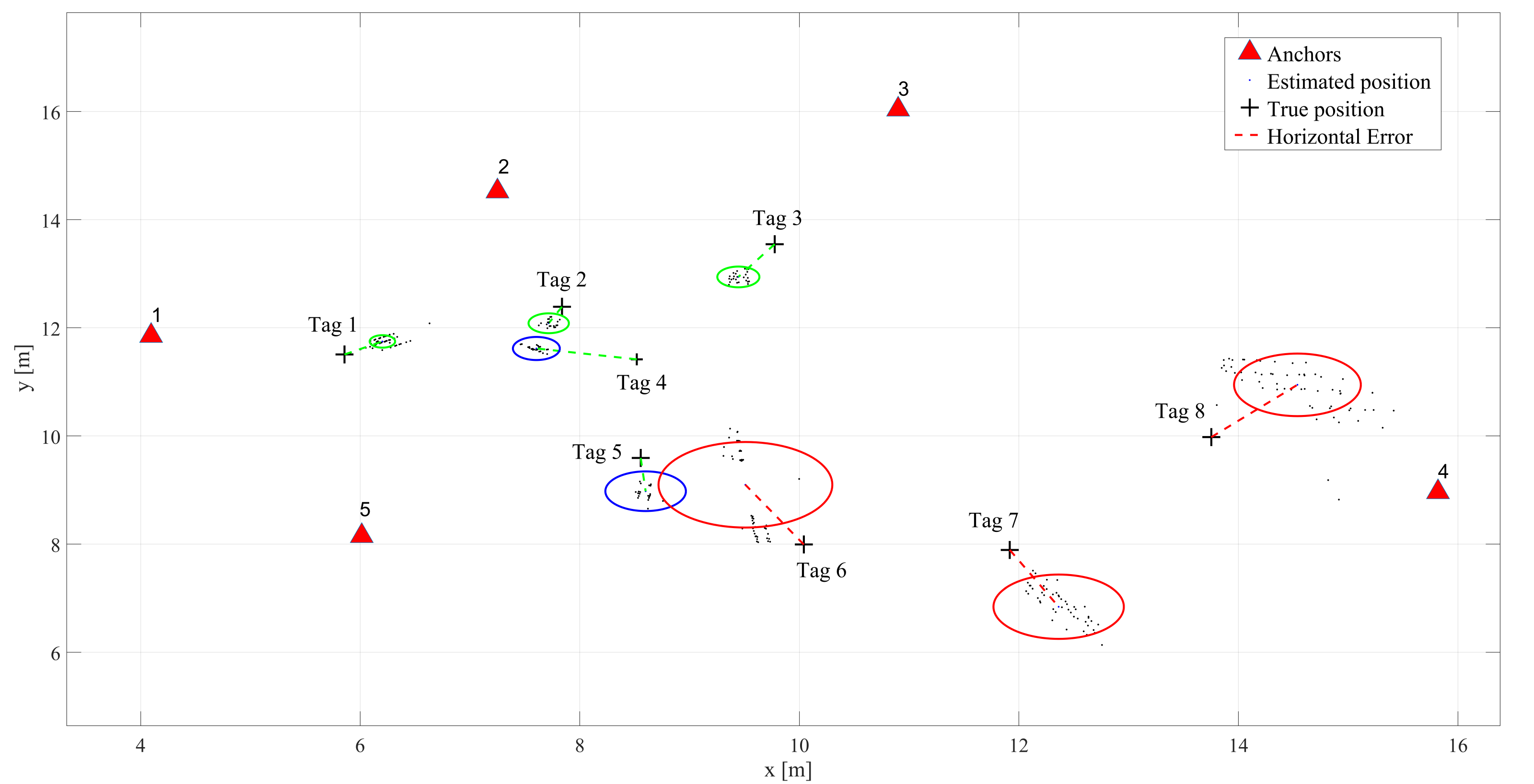}
	\caption{Positioning results for Tags 1-8 in static mode.}
	\label{fig.8}
\end{figure*}

\begin{figure*}[!t]
    \centering
	\includegraphics[width=0.85\columnwidth]{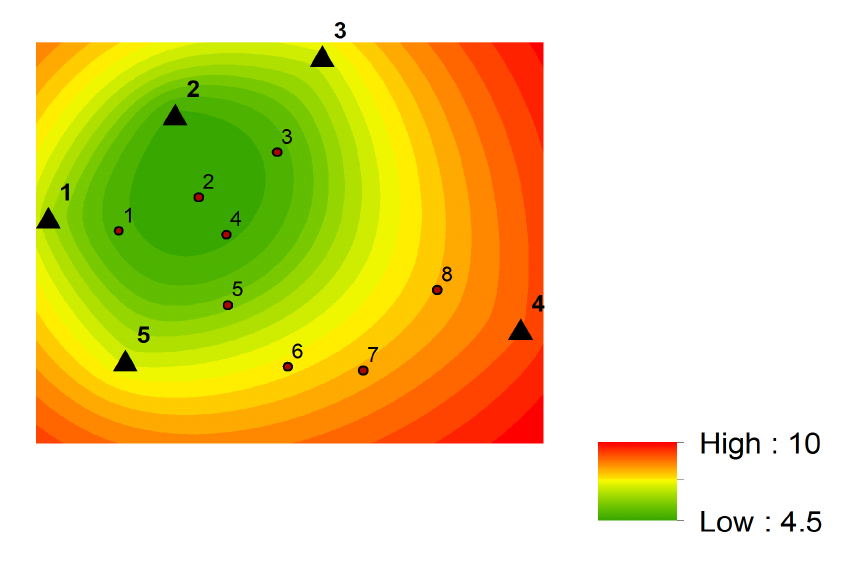}
	\caption{Map of tag positions displayed on the map of average distances (in meters) to anchors.} 
	\label{fig.9}
\end{figure*}

The tag positions were computed using the AOA positioning algorithm, which was tested in both static and kinematic modes. In the static mode, eight tags were distributed around the room, as shown in Fig.~\ref{fig.8} and Fig.~\ref{fig.9}, and observed for 1 minute. The true positions of the tags were computed using a high-precision total station with an accuracy of \(1^\circ\). 

The results show that the horizontal error of the AOA for the eight tested tags ranged from 0.33 \(m\) to 1.25 \(m\), as shown in Fig. 8. For tags 1-5, the horizontal error was less than 1 meter, attributed to stronger geometry and shorter average range compared to tags 6-9, as shown in Fig. 9.

A comparison of the results with each tag's average distance to the anchors suggests that distance to the anchors is a strong indicator of accuracy, with tags showing higher accuracy located in areas of low average distance (green area). However, this is not consistently the case; for example, tag 5 showed higher accuracy than tag 4 despite a greater average distance to the anchors, this also observed in the comparison of tags 6 and 7, and tags 6 and 8.

\begin{figure*}[!t]
    \centering
	\includegraphics[width=1\textwidth]{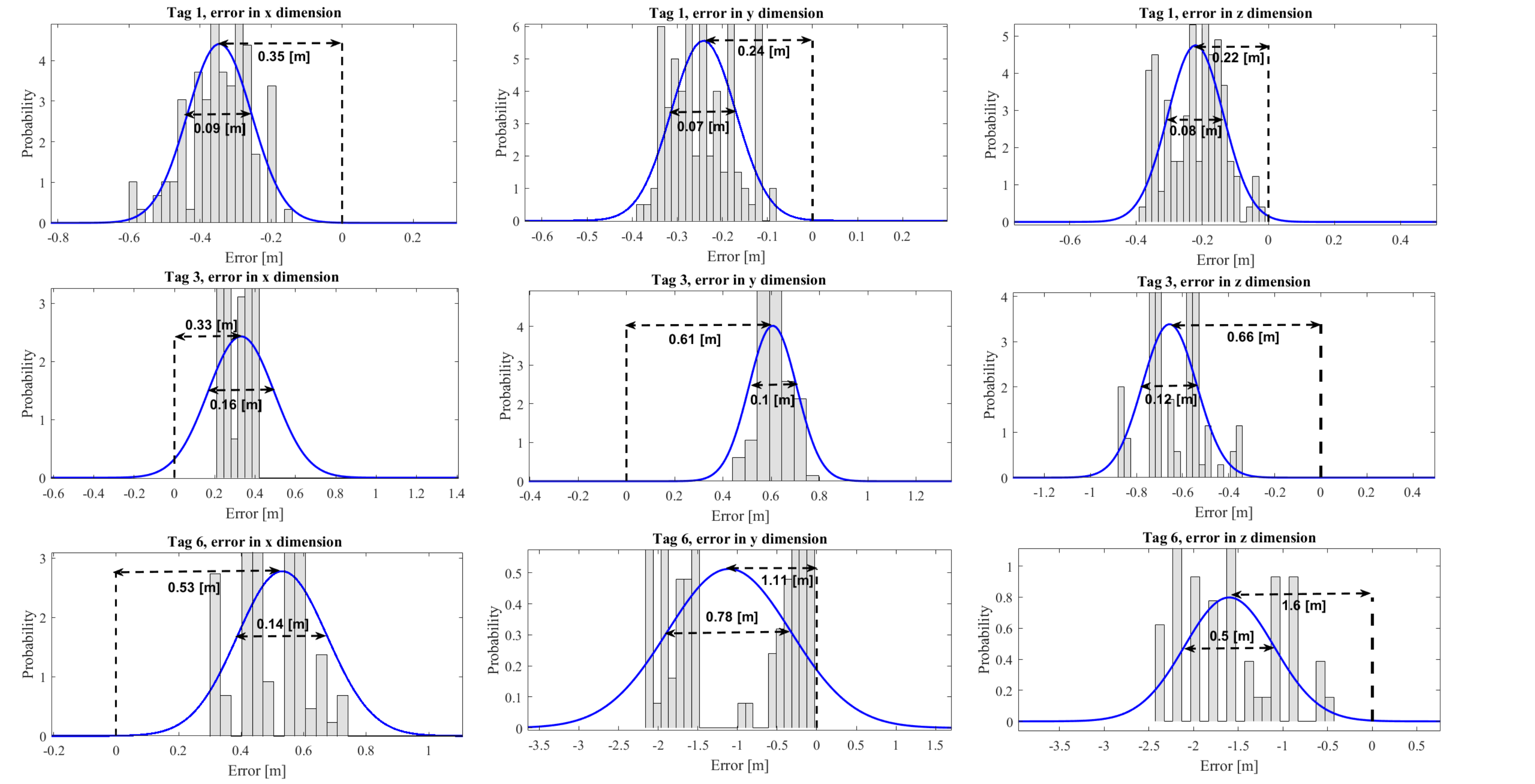}
	\caption{Error distribution in the position domain (x, y, z dimensions) for Tags 1,2, and 3} 
	\label{fig.10}
\end{figure*}

Focusing on the error distribution in the position domain (\(x, y, z\) dimensions), the results show that the error distribution does not follow a Gaussian distribution, even for those with high precision, as shown in Fig.~\ref{fig.10}. In addition, the results reveal that the impact of systematic error (bias) is much higher than that of random error, as shown in Table (2). The results show that the average impact of the systematic error is 0.44 \(m\), 0.63 \(m\), and 0.95 \(m\) in the \(x\), \(y\), and \(z\) dimensions, respectively. Meanwhile, the results show that the average standard deviation (quantifying random error) is 0.20 \(m\), 0.28 \(m\), and 0.19 \(m\) in the \(x\), \(y\), and \(z\) dimensions, respectively. By comparing the magnitude of the systematic error and random error at the 95\% confidence level, the results demonstrate that on average, the impact of random error is slightly lower than that of systematic error in the \(x\) and \(y\) dimensions. This suggests that a significant improvement, potentially doubling accuracy in the horizontal dimensions and more than doubling accuracy in the \(z\) dimension, can be achieved by developing models that effectively mitigate bias.

\begin{table*}[!t]
\caption{Analysis of Systematic and Random Error Impact on Positioning Accuracy in the \(x\), \(y\), and \(z\) Dimensions, Showing Average Values for Systematic Error (Bias) and Standard Deviation (Random Error)}
\label{table_error_analysis}
\centering
\footnotesize
\begin{tabular}{|c|c|c|c|c|}
\hline
ID & Horizontal Standard Deviation [m] & Vertical Standard Deviation [m] & Mean Horizontal Error [m] & Mean Vertical Error [m] \\
\hline
1 & 0.12 & 0.08 & 0.42 & 0.22 \\
2 & 0.18 & 0.06 & 0.33 & 0.53 \\
3 & 0.19 & 0.12 & 0.69 & 0.66 \\
4 & 0.21 & 0.05 & 0.94 & 1.15 \\
5 & 0.37 & 0.14 & 0.62 & 0.49 \\
6 & 0.79 & 0.50 & 1.23 & 1.60 \\
7 & 0.59 & 0.30 & 1.14 & 2.12 \\
8 & 0.58 & 0.28 & 1.25 & -0.87 \\
\hline
\end{tabular}
\end{table*}

\begin{figure*}[!t]
    \centering
	\includegraphics[width=0.45\textwidth]{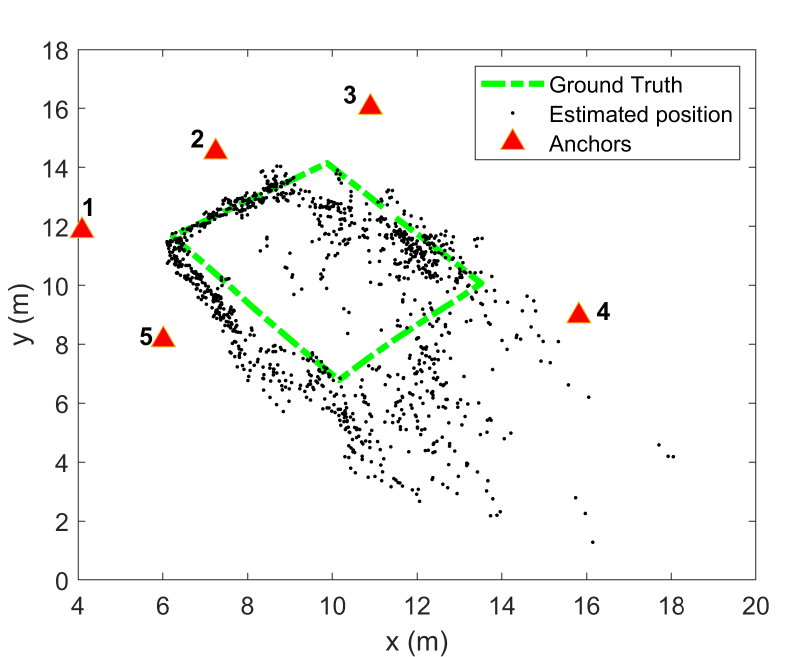}
	\caption{Kinematic mode positioning results} 
	\label{fig.11}
\end{figure*}

\begin{figure}[!t]
    \centering
	\includegraphics[width=\columnwidth]{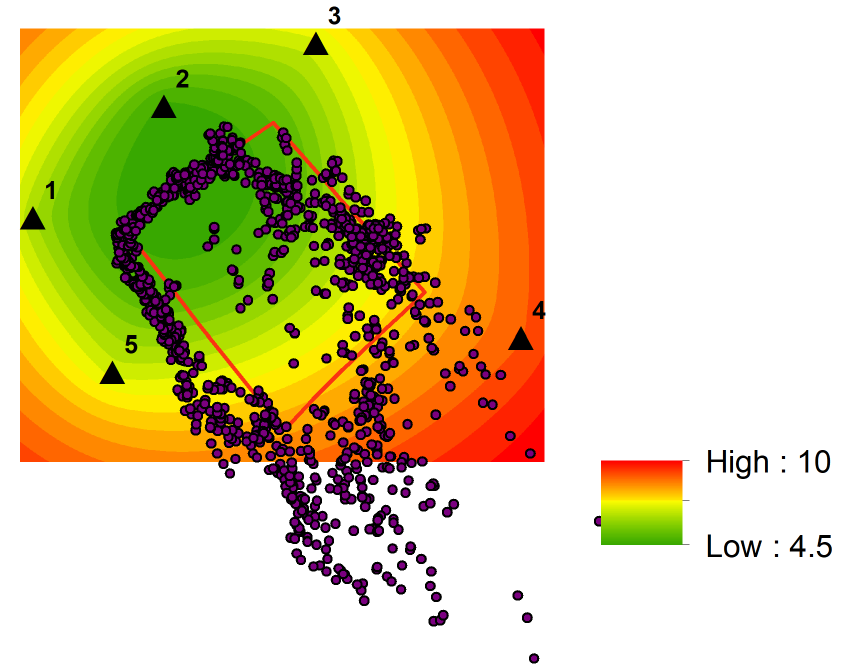}
	\caption{Kinematic mode positioning results displayed on the map of average distances (in meters) to the anchors} 
	\label{fig.12}
\end{figure}

In addition to the tests in static mode, the test was also applied in kinematic mode for pedestrians walking along a path shown in ~\ref{fig.11} and ~\ref{fig.12}. The ground truth was computed using a total station, and the across-the-path error was calculated using the computed positions and the ground truth coordinates. The results show that the across-path horizontal error average is 1.06 \(m\), with a maximum value of 8 \(m\), and the vertical accuracy is on average 1.18 \(m\), with a maximum value of 3.43 \(m\). The accuracy was highly dependent on the region, as shown in Fig. 12, for points in the green area, which have the highest accuracy.

\section{Conclusion and future work}
\label{sec:CONCLUSION AND FUTURE WORK}

To the best of our knowledge, no model or technique has yet been developed to estimate the orientations of AoA anchors required for AoA positioning. This paper introduces a novel calibration model designed to compute the orientation of AoA anchors and presents the AoA positioning model along with its functional architectures. 

The results show that the calibration models can estimate anchors' orientation in space, computing rotation around the \(x\), \(y\), and \(z\) axes with a standard deviation between \(0.44^\circ\) and \(2.19^\circ\). This accuracy could be further enhanced by incorporating more tags in the calibration mode.

The AoA positioning model has been tested in both static and kinematic modes. The results indicate that AoA positioning achieves accuracy levels ranging from decimetre to meter, influenced by the number of anchors, the distance from the anchors, and geometry.

The calibration model developed in this study can be implemented in real-world settings to calibrate AoA anchors, supporting a wide range of indoor positioning applications. This paper has focused on proving the concepts behind the calibration and AoA positioning models, paving the way for future work to enhance these models with the following roadmap:

\begin{itemize}
    \item \textbf{Development of Bluetooth AoA Network Configuration Models:} Future work should develop Bluetooth AoA configuration models to determine the optimal number, location, and orientation of tags within a target area, accounting for the installation site. This process aims to adapt the anchor network to meet system requirements. Inputs to this model include specifications for anchors and tags (e.g., measurement accuracy) and environmental characteristics, while outputs should define the number, positioning, and orientation of anchors and associated infrastructure costs.
    
    \item \textbf{Weighted Calibration and Positioning Models:} Future work should explore weighted calibration and positioning models incorporating a stochastic framework to enhance performance over the non-weighted models presented in this paper.
    
    \item \textbf{Sensitivity Analysis:} Future work should include a sensitivity analysis correlating the number of tags and their geometry with calibration model accuracy, potentially leading to a standardized calibration process.
    
    \item \textbf{Outlier Detection and Exclusion Model:} Future work should develop outlier detection and exclusion models within the positioning layer to exclude failures based on quality indicators.
    
    \item \textbf{Anchor Selection Models:} Future work should develop anchor selection models within the positioning layer. This could improve system performance by dynamically adapting anchor selection based on quality indicators (e.g., RSSI).
    
    \item \textbf{Systematic Error Mitigation Models:} Future work should develop models to mitigate systematic errors, involving further investigation into failure modes and modeling approaches.
\end{itemize}

\section{Biography Section}
\vspace{-1cm}
\begin{IEEEbiography}[{\includegraphics[width=1in,height=1.15in,clip]{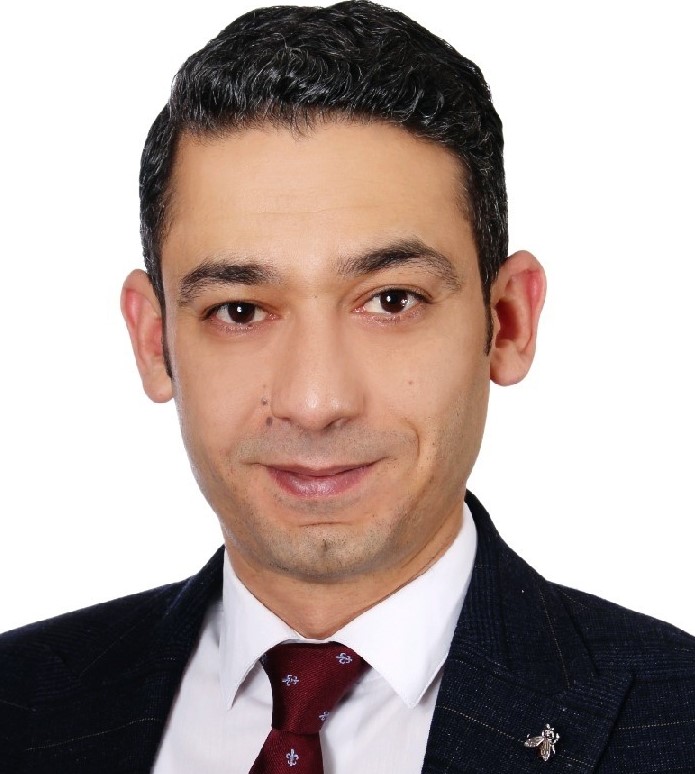}}]{Ma'mon Saeed Alghananim} received the B.S. degree in Surveying and Geomatics Engineering from Al-Balqa Applied University, Jordan, in 2011, the M.S. degree in Geospatial Engineering from the University of New South Wales, Australia, in 2016, and the Ph.D. degree in Civil and Environmental Engineering from Imperial College London, U.K., in 2022. He is currently a Research Associate at the Centre for Transport Engineering and Modelling, Imperial College London. His research interests encompass a wide range of Positioning, Navigation, and Timing (PNT) systems, including GNSS integrity monitoring, error characterization, indoor positioning, crowd-sourced positioning, and navigation systems for the visually impaired.
\end{IEEEbiography}

\vspace{-1cm}
\begin{IEEEbiography}[{\includegraphics[width=1in,height=1.15in,clip]{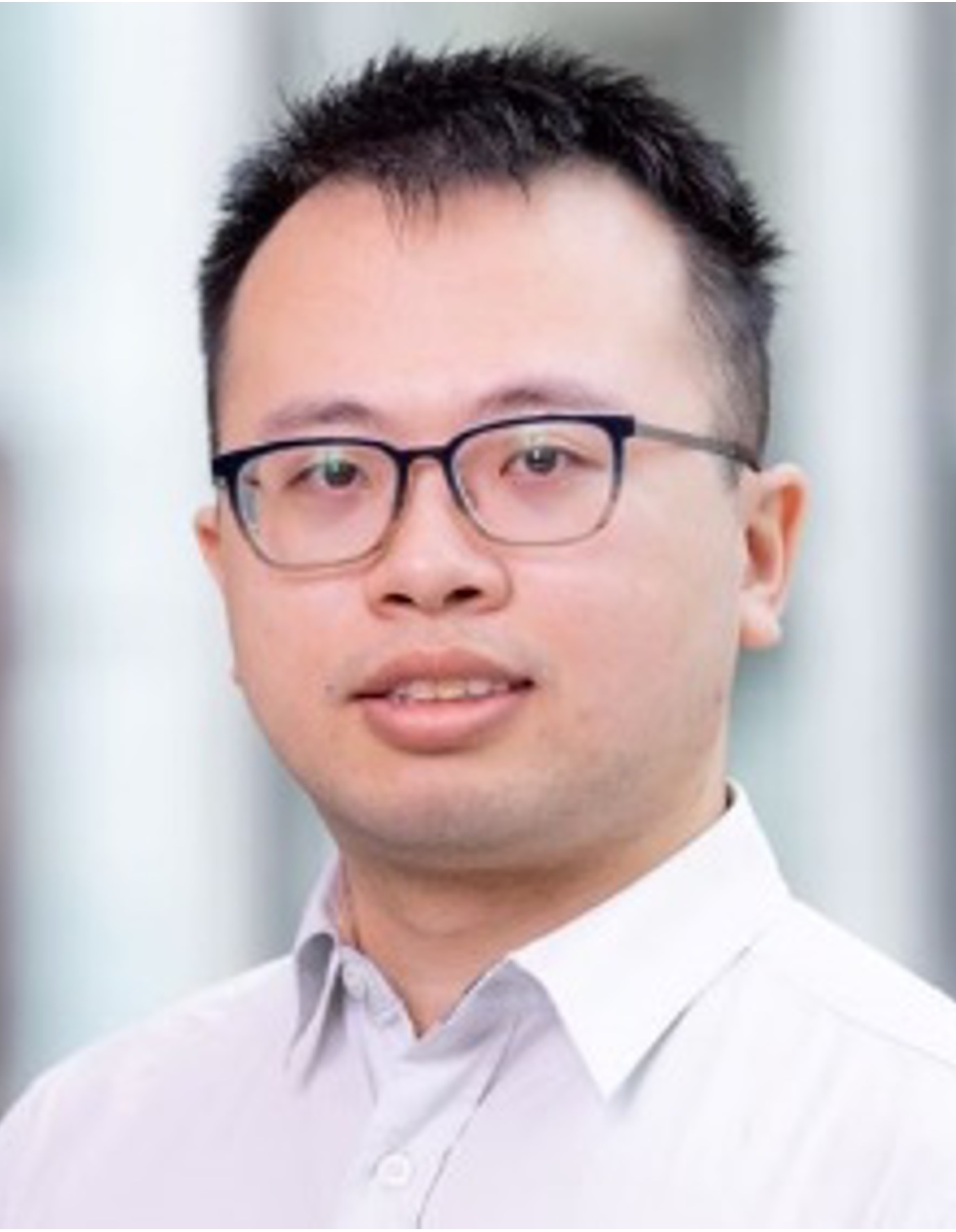}}]{Yuxiang Feng} is a Research Associate and Lab Manager of the Intelligent Infrastructure and Transport Systems Laboratory at the Centre for Transport Engineering and Modelling, Imperial College London. He received a BEng in Mechanical Engineering from Tongji University and an MSc in Mechatronics and PhD in Automotive Engineering from the University of Bath. His main research interests include environment perception, sensor fusion and artificial intelligence for robotics and autonomous vehicles.
\end{IEEEbiography}

\vspace{2cm}
\begin{IEEEbiography}[{\includegraphics[width=1in,height=1.15in,clip]{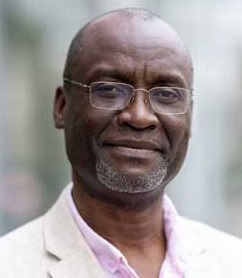}}]{Washington Yotto Ochieng} is currently a Chair Professor in positioning and navigation systems, the Head of the Department of Civil and Environmental Engineering, and the Director of the Institute for Security Science and Technology, Imperial College London, London, U.K. Professor Ochieng holds a BSc in Engineering from the University of Nairobi, Kenya, MSc and PhD degrees in Civil Engineering from the University of Nottingham, UK, and DSc (honoris causa) from the Technical University of Kenya (TUK), Kenya. He is a Chartered Engineer (CEng), Fellow of the Royal Academy of Engineering (FREng) and holds the prestigious title of Elder of the Order of Burning Spear (EBS) for service to Kenya and the world. Professor Ochieng is one of the pioneers of Europe’s space-based positioning, navigation and timing programmes. His research interests include the design of positioning and navigation systems for land, sea, and air applications and also air traffic management and intelligent transport systems.
\end{IEEEbiography}

\end{document}